%% file: paper-clean.tex
\documentclass[twocolumn,showpacs,aps,prl,superscriptaddress,nofootinbib]{revtex4}

\usepackage{graphicx}

\usepackage{dcolumn}

\usepackage{amsmath}

\usepackage{epsfig}

\input pubboard/babarsym

\def\true{{\rm true}}

\def\BDstarpi{$\Bz \rightarrow {\Dstar}^- \pi^{+}$}

\def\BDstarh{$ \Bz \rightarrow {\Dstar}^- h^{+}$}

\def\BDstarrho{${\Bz \rightarrow {\Dstar}^-  \rho^{+}}$}

\def\ifb{\rm fb^{-1}}

\def\cont{{\rm cont}}
\def\peak{{\rm peak}}
\def\DX{D^*X}
\def\comb{{\rm comb}}
\def\sig{{\rm sig}}

\def\F{{\cal F}}
\def\K{{\cal K}}
\def\M{{\cal M}}
\def\R{{\cal R}}
\def\D{{\cal D}}

\def\dz{\Delta z}

\def\Dm{\Delta m}
\def\Dt{\Delta t}

\def\tauB{\tau_{_{\hspace{-2pt}B^0}}}

\def\B{{B}}

\def\fisherpi{F_{\pi}}
\def\fisherrho{F_{\rho}}
\def\mrho{m(\pi^+\pi^0)}
\def\mmiss{m_{\rm miss}}
\def\rhoHelic{\cos\theta_{\rho}}
\def\dstHelic{\cos\theta_{\Dstar}}
\def\aOne{{\Bz \rightarrow {\Dstar} ^- a_1^+}}
\def\DstAOne{{\Dstar}a_1}

\def\BDstst{B\rightarrow D^{**} \rho^+}
\def\zrec{z_{\rm rec}}
\def\ztag{z_{\rm other}}

\def\Dt{\Delta t}

\def\DtErr{\sigma_{_{\Dt}}}

\def\mc{Monte Carlo}
\def\lumi{20.7}
\def\lumioff{2.6}
\def\nB{22.7 million}

\def\pihPi{\pi_h}

\newcommand{\SLACPubNumber} {9602}

\def\figurebox#1#2#3{%
    \def\arg{#3}%
    \ifx\arg\empty
    {\hfill\vbox{\hsize#2\hrule\hbox to #2{\vrule\hfill\vbox to #1{\hsize#2\vfill}\vrule}\hrule}\hfill}%
    \else
    {\hfill\epsfbox{#3}\hfill}%
    \fi}

\begin{document}

\preprint{\babar-PUB-02/013} 
\preprint{SLAC-PUB-\SLACPubNumber} 

\begin{flushleft}
\babar-PUB--02/013\\
SLAC-PUB-\SLACPubNumber\\
\end{flushleft}

\title{
{\large \bf Measurement of the \Bz\ Meson Lifetime with Partial
Reconstruction of \BDstarpi\ and \BDstarrho\ Decays } 
}

 \input pubboard/authors_02015.tex

\date{\today}

\begin{abstract}
The neutral $B$ meson lifetime is measured with the data
collected by the \babar\ detector at the PEP-II storage ring during
the years 1999 and 2000, with a total integrated luminosity of 20.7
$\ifb$.  The decays \BDstarpi\ and \BDstarrho\ are selected with
a partial-reconstruction technique, yielding samples of $6970 \pm 240$
and $5520 \pm 250$ signal events, respectively.  With these events, the
$B^0 $ lifetime is measured to be $ 1.533 \pm 0.034~{\rm (stat.)} \pm 0.038~{\rm (syst.)}
~{\rm ps}$.  This measurement serves as a test and validation of procedures
required to measure the \CP violation parameter 
$\sin(2\beta + \gamma)$ with partial
reconstruction of these modes.
\end{abstract}


\maketitle
\vskip .3 cm

The neutral $B$ meson decay modes~\cite{ref:ft1}
$\Bz \rightarrow {\Dstar}^- h^+$, where
$h^+$ is a light hadron ($\pi^+, \rho^+, a_1^+$), have been
proposed for use in theoretically clean measurements of 
$\sin(2\beta+\gamma)$~\cite{ref:book}, where $(2\beta+\gamma)$ is 
a combination of angles of the 
Cabibbo-Kobayashi-Maskawa~\cite{ref:km} unitarity triangle.
Since the time-dependent \CP asymmetries in these modes are expected to
be of order 2\%, large data samples and multiple decay channels are
required for a statistically significant measurement. 
The technique of partial reconstruction of $D^{*-}$ mesons, in which only 
the soft pion $\pi_s$ from the decay $D^{*-} \rightarrow \Dzb \pi_s^- $ is reconstructed, 
has already been used to select large samples of $B$ meson 
candidates~\cite{ref:cleo-dstpi}. 
This technique is applied here to the decays \BDstarpi\  and  \BDstarrho\
in order to measure the $\Bz$ lifetime.
This analysis constitutes a first step toward measuring $\sin(2\beta+\gamma)$, 
validating the procedures developed for candidate reconstruction, background characterization,
vertex reconstruction, and fitting of decay time distributions.
These procedures address 
the main complications introduced by partial reconstruction, namely
the large background and the tracks originating from the unreconstructed \Dzb,
which may affect the vertex reconstruction.

The analyses applied to the \BDstarpi\ and \BDstarrho\ modes are
similar. Detailed differences between them are the result of
optimization in the presence of the different background
characteristics in the two modes. Additional details regarding the
analysis procedures can be found in Refs.~\cite{ref:dstpiconf}
and~\cite{ref:dstrhoconf}.

The data used in this analysis were collected with the \babar\
detector at the \pep2\  asymmetric-energy storage ring during
the years 1999 and 2000. The data consist of \nB\ $\BB$
pairs, corresponding to an integrated luminosity of \lumi~$\invfb$
recorded at the $\FourS$ resonance.  In addition, \lumioff~$\invfb$
of ``off-resonance'' data were collected about 40~MeV below the resonance.
Samples of simulated \BB and continuum $\epem \rightarrow \qqbar$
events, where $q$ stands for a $u$, $d$, $s$, or $c$ quark, were
generated using a GEANT3-based detector simulation~\cite{ref:geant}
and processed through the same reconstruction and analysis chain as
the data.
The equivalent luminosity of the simulated events is approximately
one third the data luminosity. We also used signal Monte Carlo samples
with an equivalent luminosity several times
larger than that of the data.

The \babar\ detector, described in detail elsewhere~\cite{ref:nim},
consists of five subdetectors.  Charged particle trajectories are
measured by a combination of a five-layer silicon vertex tracker (SVT)
and a 40-layer drift chamber (DCH) in a 1.5~T solenoidal magnetic field.
Tracks with low transverse momentum are reconstructed by the SVT
alone, thus extending the charged particle detection down to
transverse momenta of $\sim 50$\mevc. Photons and electrons are
detected in a CsI(Tl) electromagnetic calorimeter (EMC), with 
photon energy resolution $\sigma_E / E = 0.023 (E/\gev)^{-1/4} \oplus 0.019$. A
ring-imaging Cherenkov detector (DIRC) is used for charged particle
identification. The instrumented flux return (IFR) is equipped with
resistive plate chambers to identify muons.

In the partial reconstruction of a \BDstarh\ candidate, 
only the hadron $h$ and the $\pi_s$
tracks are reconstructed. The angle between the momenta of the $B$ and
the $h$ in the center-of-mass (CM) frame is then computed:
\begin{equation}
\cos\theta_{Bh} = 
        {M_{{\Dstar}^-}^2 - M_{\Bz}^2 - M_{h}^2 + E_{\rm CM} E_h
                        \over
         2 p_B |\vec p_{h}|
        },
\label{eq:cosTheta}
\end{equation}
where $M_x$ is the mass of particle~$x$, $E_h$ and $\vec p_h$
are the measured CM energy and momentum of the hadron $h$, $E_{\rm CM}$ is
the total CM energy of the beams, and 
$p_B = \sqrt{E_{\rm CM}^2/4 - M_{\Bz}^2}$. All masses refer to the nominal
values~\cite{ref:pdg}, except in the case $h=\rho$, where the measured
$\pi^+ \pi^0$ invariant mass $\mrho$ is used.
Events are required to be in the 
physical region  $|\cos\theta_{Bh}|<1$.
Given $\cos\theta_{Bh}$ and the measured four-momentum of $h$, 
the $B$ four-momentum can be calculated up to
an unknown azimuthal angle $\phi$ around ${\vec p}_{h}$. For every
value of $\phi$, the expected $\Dzb$ four-momentum ${\cal
P}_D(\phi)$ is determined from four-momentum conservation, and the
$\phi$-dependent ``missing mass'' is calculated,
$m(\phi) \equiv \sqrt{|{\cal P}_D(\phi)|^2}$.
We define the missing mass
$\mmiss \equiv {1 \over 2}\left[m_{\rm max} + m_{\rm min}\right]$,
where  $m_{\rm max}$ and $m_{\rm min}$ are the maximum and minimum values
of $m(\phi)$.
In signal events, this variable peaks at the nominal $\Dzb$ mass
$M_{\Dz}$, with a spread of about 3 \mevcc for \BDstarpi\ (3.5 \mevcc for
\BDstarrho)~\cite{ref:ft2}, while the distribution of background
events is broader. The missing mass is the main variable used to 
distinguish signal from background.

We define the $\Dstar$ helicity angle $\theta_{\Dstar}$ to be the
angle between the directions of the $\Dzb$ and the $\Bz$ in the $\Dstar$
rest frame. This variable is used in the event selection described below.
In the \BDstarpi\ analysis, $\theta_{\Dstar}$ is computed assuming
that the \B\ momentum lies in the plane defined by the $h$ and $\pi_s$
momenta in the CM frame. This assumption also yields the $\Dzb$ direction.  
In the \BDstarrho\ analysis, the value of
$\dstHelic$ is computed by applying the constraint 
$\mmiss = M_{\Dz}$ giving two possible solutions for the $\Dzb$ direction \cite{ref:cleo-dstpi}.
In \BDstarrho, the $\rho$ helicity angle $\theta_{\rho}$ is defined as
the angle between the directions of the $\piz$ (from the decay of the
$\rho$) and the CM system in the $\rho$ rest frame.

We select events in which the ratio of the 2nd to the 0th Fox-Wolfram
moment~\cite{ref:R2}, computed using charged particles, is smaller
than 0.35. 
The candidate $\Bz$ daughter tracks are required to 
originate within 1~cm (1.5 cm)  of the 
interaction point in the $x$-$y$ plane (the plane
perpendicular to the beams), and within $\pm 4$~cm ($\pm 10$~cm) 
of the interaction point along the direction of the beams.
Tracks are rejected 
if they are highly likely to be a kaon or a lepton on the basis
of their ionization, Cherenkov angle, energy deposited in 
the EMC, and pattern of hits in the IFR.  

\BDstarpi\ candidates are rejected if another track is found within
0.4 rad of the momentum of the hard pion $\pihPi$~\cite{ref:ft3} in the CM
frame. This requirement helps to reject continuum
events, where tracks tend to be clustered in jets.  A Fisher
discriminant \cite{ref:fisher} $\fisherpi$ is computed from 15 event shape variables. Among
these variables is the scalar sum of the CM momenta of all tracks and
neutral candidates in nine $20^{\circ}$ single-sided cones around the
$\pihPi$ direction.  
We require $| \cos \theta_{\Dstar} |$ to be larger than
0.4.
A cut on $\fisherpi$ is used to reduce the continuum
background. 

In the reconstruction of \BDstarrho\ candidates, the charged $\rho$
candidates are identified by their decay to a hard charged pion
$\pihPi$ and a $\piz$. To suppress fake $\piz$ candidates, the $\piz$
momentum in the CM frame is required to be greater than 400\mevc . The
invariant mass of the $\piz \to \gamma \gamma$  candidate 
must be within 20\mevcc of the
nominal $\piz$ mass~\cite{ref:pdg}.
The invariant mass $\mrho$ of the $\rho$  candidate must be
between 0.45 and 1.10~\gevcc.
To suppress combinatoric background, we require $|\rhoHelic| > 0.3$
and $|\dstHelic|>0.3$, and also reject events that satisfy both $\rhoHelic >
0.3$ and $\dstHelic<-0.3$.
A Fisher discriminant $\fisherrho$ is computed using the scalar sum of
the CM momenta of all tracks and neutrals in nine $10^{\circ}$
double-sided cones around the $\rho$ direction.
In about 10\% of the events, more than one partially reconstructed
candidate per event satisfies all the requirements. In such events
only the candidate with the smallest value of $|\mmiss - M_{\Dz}|$ in
the event is used.

The decay position $\zrec$ of the partially reconstructed $B$
candidate along the beam direction is determined by constraining the
$\pihPi$ and the $\pi_s$ tracks (only the $\pihPi$ track for
\BDstarrho) to originate from the beam-spot in the $x$-$y$ plane.  The
beam spot is determined on a run-by-run basis using two-prong events
\cite{ref:nim}. Its size in the horizontal direction is 120 $\mu$m.
Although the beam spot size in the vertical direction is only a few
microns, a beam spot constraint of 30 $\mu$m is applied, so as to
account for the flight of the $B^0$ in the vertical direction.

The decay position $\ztag$ of the other $B$ meson along the beam
direction is measured with all tracks, excluding $\pihPi$,  $\pi_s$,
and any track whose CM angle with respect to the $\Dzb $ direction
(either of the two calculated directions in the \BDstarrho\ case) is
smaller than 1~radian. This ``cone cut'' reduces significantly
the number of $\Dzb$ daughter tracks used in the other $B$
vertex. The tracks satisfying this requirement are fit with a
constraint to the beam-spot in the $x-y$ plane.  The track with the
largest contribution to the $\chi^2$ of the vertex, if greater than 6,
is removed from the vertex, and the fit is carried out again, until no
track fails this requirement. 
\BDstarpi\ candidates are required to have at least two tracks remaining in
the other B vertex.  

The $z$ distance between the two $B$ decay vertices, $\Delta
z~=~\zrec~-~\ztag$, is computed. Fitting the residual $\Delta z - \Delta
z_{\true}$ in simulated events, where $ \Delta z_{\true}$ is the true $ \Delta
z $,  with the sum of two  Gaussians, we find that 67\% (57\%) of
the \BDstarpi\ (\BDstarrho) events lie in the core Gaussian of width
$116~\mu$m ($178~\mu$m).
The $\Delta z$ resolution is dominated by the
measurement of $\ztag$, and by the $\zrec$ measurement
when the $\pihPi$ transverse momentum is below about 400~\mevc.

The decay time difference $\Dt $ is then calculated using the
approximation $\Dt \approx \Delta z / (\gamma\beta c)$, where the CM
frame boost $\gamma\beta$ is determined from the beam energies, and has
an average value of 0.55. This approximation results in a 0.2~ps r.m.s.~spread 
in the calculation of $\Dt$.

For \BDstarpi\ candidates, $\Dt$  is computed
applying an event-by-event correction to the measured value
of $\Delta z $. This correction, determined from the simulated signal
sample as a function of $\Delta z $, removes the bias  
in $\ztag$ due to the tracks coming from the $\Dzb$ decay. Without correction, 
the effect of this bias would be to reduce the measured lifetime by approximately 4\%.
In the \BDstarrho\ analysis a different correction is applied to the measured
lifetime value, as explained later.

The estimated error $\DtErr$ in the measurement of $\Dt$ is calculated
from the uncertainties in the parameters of the tracks used 
in the two vertex fits.  
A requirement on the vertex fit probabilities removes badly
reconstructed vertices.
For both modes we also require $|\Dt| < 15$~ps
and $ \DtErr < 2.4$~ps ($ \DtErr < 4$~ps for \BDstarrho).

After applying all the above requirements, we find four broadly-defined types
of events that contribute to the background: (1) continuum events; (2)
combinatoric \BB  background due to random $h$ and $\pi_s$
combinations; (3) \BDstarrho\ ($\aOne$) decays in the \BDstarpi\
(\BDstarrho) sample; (4) peaking \BB  events, which are
distributed as a broad peak in the $\mmiss$ spectrum. The peaking
background is mostly due to $B \rightarrow D^{**} \pi $ decays in
the \BDstarpi sample. In the \BDstarrho\ sample, it is due to signal events in which the
$\pihPi$ candidate originates from the other $B$.

The lifetime $\tau_{B^0}$ is obtained from an unbinned maximum likelihood
fit, as described below, with a probability density function (PDF) 
$ {\F}(\Dt, \DtErr, \xi)$. Here $\xi$ refers to the 
kinematic variables used to distinguish signal from background.
For \BDstarpi\ we set $\xi = \mmiss$; for \BDstarrho\ we set  
$\xi = (\mmiss,\mrho, \fisherrho)$.
The PDF has the form
\begin{eqnarray}
\F (\xi,\Dt, \DtErr) &=& 
     f_{\cont}  \K_{\cont} (\xi) \F_{\cont} (\Dt, \DtErr) \nonumber\\
 &+& f_{\comb}  \K_{\comb} (\xi) \F_{\comb} (\Dt, \DtErr)  \nonumber\\
 &+& f_{\DX}    \K_{\DX} (\xi)   \F_{\DX} (\Dt, \DtErr)  \nonumber\\
 &+& f_{\peak}  \K_{\peak} (\xi) \F_{\peak} (\Dt, \DtErr)  \nonumber\\
 &+& f_{\sig}   \K_{\sig} (\xi)  \F_{\sig} (\Dt, \DtErr), 
\label{eq:pdf}
\end{eqnarray}
where subscripts 
refer to the four types of backgrounds enumerated above and to signal
events. For each event type $i$, $f_i$ is the relative population
of these events in the data sample, $\K_i(\xi)$ is their 
kinematic-variables PDF, and $\F_i(\Dt, \DtErr)$ is their time-dependent PDF. 
The constraint $\sum f_i = 1$ is enforced.

For \BDstarpi, $\K_i (\mmiss)$ consists of binned hi\-sto\-grams obtained from
the Monte Carlo simulation. For \BDstarrho\ candidates, we use the
product $\K_i (\xi) = \M_i(\mmiss) \R_i(\mrho) \D_i(\fisherrho)$, where
$\M_i(\mmiss)$ is the sum of a bifurcated Gaussian and an ARGUS
function~\cite{ref:argus}, $\R_i(\mrho)$ is the sum of a parabolic
background and a relativistic $P$-wave Breit-Wigner, 
and $\D_i(\fisherrho)$ is a bifurcated Gaussian function.

For each event type $i$, $\F_i(\Dt, \DtErr)$ is the  convolution
$N \int
        P(\Dt_{\true}) R((\Dt - \Dt_{\true})/ \DtErr) d\Dt_{\true}$
of the ``true'' distribution $P(\Dt_{\true})$ and the detector
resolution function $R((\Dt - \Dt_{\true})/ \DtErr)$, which is
parameterized as the sum of three Gaussian distributions. 
$N$ is a normalization constant.
The parameters of $P(\Dt_{\true})$ and $R((\Dt - \Dt_{\true})/
\DtErr)$ are obtained separately for each event type.
For signal events of both modes we take $P(\Dt_{\true}) =  
e^{-|\Dt_{\true}| / \tauB}$.
This functional form is also used for the combinatoric and
peaking backgrounds in \BDstarpi, but with independent parameters. 
In \BDstarrho, the source  of
the peaking background motivates its distribution to be 
$P(\Dt_{\true}) = \delta(\Dt_{\true})$, 
and the distribution used for the combinatoric background is
$P(\Dt_{\true}) = a e^{-|\Dt_{\true}| / \tau'} + 
(1 - a) \delta(\Dt_{\true}) $, 
with an effective lifetime parameter $\tau'$.
$\F_{\DX}(\Dt, \DtErr)$ is assumed to be identical to  
$\F_{\sig}(\Dt, \DtErr)$.
The continuum background is modelled as
$P(\Dt_{\true}) = b e^{-|\Dt_{\true}| / \tau_{\cont}} + 
(1 - b) \delta(\Dt_{\true})$.

Several subsamples are defined and used in the lifetime fit. 
Events with a candidate in which the $h$ and $\pi_s$ have opposite
charges and with $\mmiss > 1.860 $\gevcc ($\mmiss~>~1.845 \gevcc$ 
in \BDstarrho) 
constitute the ``signal region'' sample.  Those 
satisfying $1.820 < \mmiss < 1.850$\gevcc ($1.810 < \mmiss < 1.840\gevcc$)
constitute the ``sideband''.
Events in which $h$ and $\pi_s$ have the same
charge are labeled as ``same-charge''. 
In the \BDstarpi\ analysis, we apply a requirement on the Fisher
discriminant that suppresses $\BB$ events, to select a
``$\BB$-depleted'' sample that is enriched in continuum events.
The sideband, same-charge, and $\BB$-depleted samples serve as control
samples for studying the $\Dt$ distributions of the backgrounds.

In the \BDstarrho analysis, about 11.5\% of the partially reconstructed
signal events are also fully reconstructed in the $\Dzb$ decay modes $\Dzb
\rightarrow K^+ \pi^-$ or $K^+ \pi^- \piz$. This yields a sample that,
while relatively small, has a low background contamination of about 5\%. 
This clean signal sample is used in the fits described below, improving
the determination of the signal PDF parameters.

The $\Bz$ lifetime $\tauB$ is obtained in a three-step procedure
using signal region and control sample events.

\begin{figure}[!htb]
\begin{center}
\includegraphics[width=6.5cm]{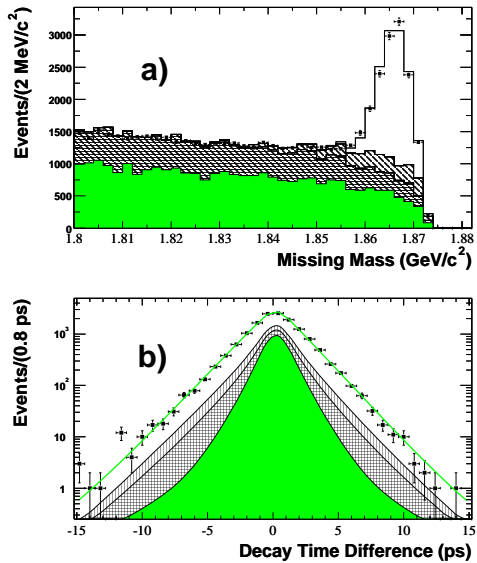}
\caption{
Distributions of (a) missing mass and (b) $\Delta t$ for candidate
\BDstarpi\ events. The result of the fit (solid line) 
is superimposed on data (data points). The
 hatched, cross-hatched and shaded areas are the peaking \BB,
combinatoric \BB, and continuum contributions, respectively.
The $\Dt$ plot is obtained with the requirement $\mmiss > 1.860 $\gevcc.}
\label{fig:dstpi}
\end{center}
\end{figure}

\begin{figure}[!htb]
\begin{center}
\includegraphics[width=9cm]{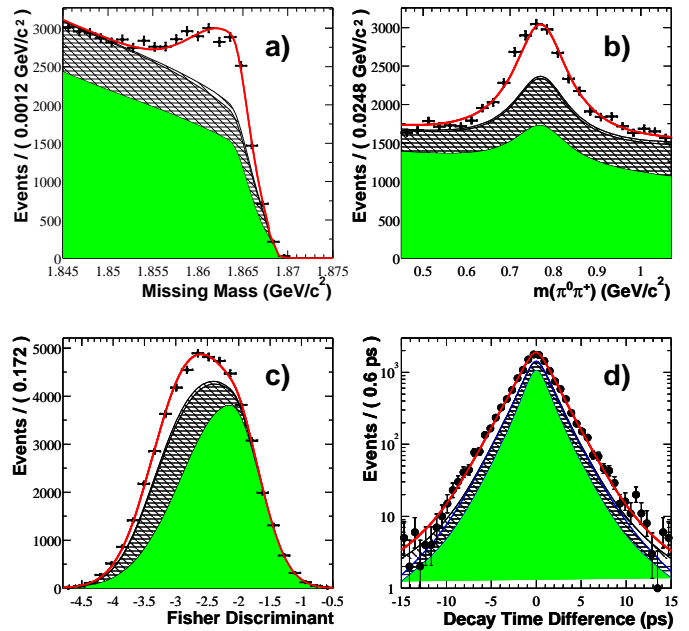}
\caption{
Distributions of (a) missing mass,
(b) $\rho$ candidate invariant mass, 
(c) Fisher discriminant $\fisherrho$ 
and (d) $\Delta t$ of \BDstarrho\ candidate events. 
The result of the fit (solid line) is
superimposed on data (data points). 
The hatched, cross-hatched and shaded areas are the
peaking \BB, combinatoric \BB, and continuum
contributions, respectively. The $\Dt$ plot is obtained with the 
requirement 
$\mmiss > 1.854$\gevcc, 
$0.60 < \mrho<0.93$\gevcc, and
$\fisherrho < -2.1$.}
\label{fig:dstrho}
\end{center}
\end{figure}

In the first step, the fractions $f_i$ in the signal region and in the
different control samples are obtained from kinematic-variable fits
conducted simultaneously on the on- and off-resonance samples (and the
fully reconstructed sample for the \BDstarrho\ signal region). The fit
PDF is that of Eq.~(\ref{eq:pdf}), but with all $\F_i (\Dt, \DtErr)$
replaced by unity.
In the \BDstarpi\ analysis this fit determines $f_{\peak}$ and $f_{\cont}$.
The fraction of \BDstarrho\ events $f_{\DX}$ in the
\BDstarpi\ sample is assumed to be 16.8\%, as predicted by the \mc\
simulation and the relative branching ratio~\cite{ref:pdg} .
This fit (Fig.~\ref{fig:dstpi}(a)) yields $6970 \pm 240$ signal
\BDstarpi\ events.
In the \BDstarrho\ analysis the kinematic-variable fit determines $f_{\cont}$,
as well as all the parameters of $\K_{\cont}(\xi)$, $\M_{\sig}(\mmiss)$,
and $\R_{\sig}(\mrho)$. The parameters of $\D_{\sig}(\fisherrho)$,
$\K_{\comb}(\xi)$, and $\K_{\peak}(\xi)$, as well as $f_{\peak}/f_{\sig}$ 
(9.7\%) and $f_{\DstAOne}/f_{\sig}$
(11.6\%), are obtained from the \mc\ simulation.
The kinematic-variable fit to the \BDstarrho\ sample
(Figs.~\ref{fig:dstrho}(a), (b) and~(c)) yields $5520 \pm 250$
\BDstarrho\ events, including $691 \pm 36$ fully reconstructed
events. 

In the second step, all the parameters determined in the first step
are fixed, and the parameters of $\F_i(\Dt, \DtErr)$ of the backgrounds are
determined entirely from the control data samples. In the \BDstarpi\ case, the
parameters of $\F_{\rm cont}(\Dt, \DtErr)$ 
are obtained from a fit to the \BB-depleted sample, and
those of the $\F_{\rm comb}(\Dt, \DtErr)$ 
are obtained from the same-charge sample. The parameters of
$\F_{\rm peak}(\Dt, \DtErr)$ 
are assumed to be identical to $\F_{\rm comb}(\Dt, \DtErr)$.
In \BDstarrho, the parameters of $\F_{\rm comb}(\Dt, \DtErr)$ are
determined from the sideband sample, and those of $\F_{\rm peak}(\Dt, \DtErr)$
are obtained from the same-charge sample. Each of the \BDstarrho\
control sample fits is conducted simultaneously on the on- and
off-resonance data, and the parameters of $\F_{\rm cont}(\Dt, \DtErr)$ are
determined for each control sample simultaneously with the \BB PDF
parameters.

In the final step, using the background $\F_i(\Dt, \DtErr)$ parameters obtained
in the previous step, the signal region sample is fit to extract the
signal $\F_{\sig}(\Dt, \DtErr)$  parameters. 
In \BDstarpi\ this fit has six free parameters describing $\F_{\sig}(\Dt, \DtErr)$.
In \BDstarrho, the fit is done simultaneously to on- and
off-resonance events, as well as fully reconstructed events,
and has 15 free parameters describing $\F_{\sig}(\Dt, \DtErr)$ and $\F_{\cont}(\Dt, \DtErr)$.

The results of the last fit step, shown in Figs.~\ref{fig:dstpi}(b) 
and~\ref{fig:dstrho}(d), are $\tauB~=~1.510~\pm~0.040~$ps for \BDstarpi\ and
$\tauB = 1.616 \pm 0.064$~ps for \BDstarrho, where the errors are
statistical only. These results are obtained after a correction of
$-0.014 \pm 0.020$~ps ($+0.071 \pm 0.028$~ps for \BDstarrho), determined
from the Monte Carlo simulation. The correction accounts for biases due to the
fit procedure, the event selection and, in the \BDstarrho\ case, the
effect of $\Dzb$ daughter tracks passing the cone cut and being used for
the determination of the other $B$ vertex. The errors in the corrections
are propagated to the final result as systematic errors.

The systematic uncertainties are listed in Table~\ref{tab:syst}, and
described here. 
(1) The fractions and the PDF parameters of the background components
were varied by their statistical errors, taking into account mutual
correlations, obtained from the fits of the first two analysis steps.
(2) The PDF parameters and lifetime corrections that were obtained
from the \mc\ simulation were varied by the statistical error in the
\mc\ fits. The full analysis chain, including event reconstruction and
selection, was tested with the Monte Carlo simulation, and the
statistical precision of the consistency between the generated and fit
lifetimes was assigned as a systematic error. The Monte Carlo statistical
errors in the evaluation of the various corrections described above
were propagated to the final result.
(3) The level of \BDstarrho\ ($\aOne$) background in the \BDstarpi\
(\BDstarrho) sample was varied by the relevant branching fraction
errors~\cite{ref:pdg}, and the fraction of $\BDstst$ background events
in the \BDstarrho\ sample was varied between 0 and 40\%
of the signal yield.
(4) The number of $\Dzb$ tracks satisfying the cone cut in the simulated
sample was varied by $\pm 5\%$ and the associated bias was reevaluated.
(5) The parameters of $\F_{\sig}$ 
that were fixed in the fits were varied within conservative ranges.
(6) Extensive
parameterized \mc\ simulation studies were conducted to evaluate statistical
biases in the fits due to limited data sample size or as the result of
changes in the functional form of $R((\Dt - \Dt_{\true})/ \DtErr)$.
(7) The $\Dt$ fit range was varied between $|\Delta t| < 10$~ps
and $|\Delta t| < 20$~ps.
(8) The $z$~length scale of the detector has been determined with an
uncertainty of 0.4\% from the reconstruction of secondary interactions
with a beam pipe section of known length~\cite{ref:sin2b-prd}.  The
systematic uncertainties related to the detector alignment (9) and
beam energy uncertainty~\cite{ref:nim} (10) were also taken into account.
The total systematic error in the \BDstarpi\ (\BDstarrho) analysis
is 0.041~ps (0.075~ps).

\begin{table}[tbp]

\caption{ Summary of the systematic uncertainties on the measured
     $\Bz$ lifetime.  }
\begin{center}
\begin{tabular}{lcc}
\hline \hline 
\multicolumn{1}{c}{Source} &  \multicolumn{2}{c}{Errors (ps)} \\
   & \BDstarpi\ & \BDstarrho\ \\
 \hline
(1) {Background parameters}      &  0.023  & 0.044 \\
(2) {\mc\ statistics}             &  0.021  & 0.042 \\
(3) {Fractional composition}      &  0.008  & 0.024 \\
(4) {$\Dz$ tracks bias} & 0.017 & 0.026 \\
(5) {$\Delta t$ resolution model } & 0.011 & 0.015 \\
(6) {Likelihood fit bias} & 0.005 & 0.016 \\
(7){$\Delta t$ range} & 0.009 & 0.009 \\
(8) {$z$ scale} & 0.006 & 0.007 \\
(9) {SVT misalignment } & 0.008 & 0.008 \\
(10) {Beam energies } & 0.002 & 0.002 \\
\hline
{Total} & 0.041 & 0.075 \\
\hline\hline 
\end{tabular}
\end{center}
 {\label{tab:syst}}
\end{table}

Several cross-checks were conducted to ensure the validity of the
result.  
The data were fit in bins of the lab frame polar angle, azimuthal
angle, and momentum of the $\pihPi$, and in subsamples corresponding
to different SVT alignment calibrations. The fit was repeated with different values of the cone cut ranging from 0.75 to 2.00 radians (0.6 to 1.2 radians
for \BDstarrho).
Different functional forms
of $R((\Dt - \Dt_{\true})/ \DtErr)$ were used in the fit.  In all cases, no
statistically significant variation of the result was observed, beyond
those already accounted for in the systematic errors. 

\label{sec:Summary}

In summary, in a sample of \nB\ $\BB$ pairs, we identify $6970 \pm 240$   \BDstarpi\ and $5520\pm 250 $
\BDstarrho\  partially reconstructed  decays.  
These events are used to measure the $\Bz$ lifetime,
obtaining $ \tauB = 1.510 \pm 0.040~{\rm (stat.)} 
\pm 0.041~{\rm (syst.)}~\ps$ 
in \BDstarpi\ and 
$ \tauB = 1.616 \pm 0.064~{\rm (stat.)} \pm 0.075~{\rm (syst.)~ps}$
in \BDstarrho ~. 
The combined measurement, taking into account correlated errors, is 
$$ \tauB = 1.533 \pm 0.034~{\rm (stat.)} \pm 0.038{\rm (syst.)}~\ps.$$
This result is in good agreement with the world average $B^0$ lifetime 
$ \tauB = 1.542 \pm 0.016~\ps $ \cite{ref:pdg} and with other recent \babar\
measurements \cite{bbrtaub0}, confirming the validity of using 
partially reconstructed  events in time dependent measurements.

\input pubboard/acknow_PRL.tex

\end{document}

%% file: pubboard/authors_02015.tex
%
\author{B.~Aubert}
\author{R.~Barate}
\author{D.~Boutigny}
\author{J.-M.~Gaillard}
\author{A.~Hicheur}
\author{Y.~Karyotakis}
\author{J.~P.~Lees}
\author{P.~Robbe}
\author{V.~Tisserand}
\author{A.~Zghiche}
\affiliation{Laboratoire de Physique des Particules, F-74941 Annecy-le-Vieux, France }
% ...
\collaboration{The \babar\ Collaboration}
\noaffiliation

%% file: pubboard/acknow_PRL.tex
We are grateful for the excellent luminosity and machine conditions
provided by our \pep2\ colleagues, 
and for the substantial dedicated effort from
the computing organizations that support \babar.
The collaborating institutions wish to thank 
SLAC for its support and kind hospitality. 
This work is supported by
DOE
and NSF (USA),
NSERC (Canada),
IHEP (China),
CEA and
CNRS-IN2P3
(France),
BMBF and DFG
(Germany),
INFN (Italy),
NFR (Norway),
MIST (Russia), and
PPARC (United Kingdom). 
Individuals have received support from the 
A.~P.~Sloan Foundation, 
Research Corporation,
and Alexander von Humboldt Foundation.